\documentclass[aps,prb,twocolumn,epsfig,superscriptaddress,floatfix]{revtex4}

\newcommand{\dir}{Figs}
\makeatletter
\def\thepage{\@arabic\c@page}
\def\@pnumwidth{2em}

\makeatother
\usepackage[dvips]{graphicx}
\usepackage{epsfig}
\usepackage{amssymb}

\newcounter{Figure}
\newenvironment{FigureCaptions}{\begin{list}{
  Fig. \theFigure. \rm}
  {\protect\usecounter{Figure}\setlength{\labelwidth}{9em}}
  }{\end{list}}

\newcounter{Table}
\newenvironment{Table}[1]{\refstepcounter{Table}
   \begin{center} Table \theTable. \\  \rm #1 \\[2ex]
  }{\end{center}}

\newcommand{\eref}[1]{Eq. \ref{eqn:#1}}

\newcommand{\fig}[2]
{
\noindent
Fig. #2 
Cheung and Schmid \\
Journal of Chemical Physics
\\
\\
\includegraphics[width=7cm]{#1}

\vfill

\clearpage
}

\begin{document}

\title{
Monte Carlo simulations of liquid crystals near rough walls
}

\author{David L. Cheung
\footnote{Present address: Dept. of Physics, University of Warwick,
  Coventry, CV4 7AL, UK}
}
\affiliation{Theoretische Physik, Universit\"{a}t Bielefeld,
33615 Bielefeld, Germany}
\author{Friederike Schmid}
\affiliation{Theoretische Physik, Universit\"{a}t Bielefeld,
33615 Bielefeld, Germany}

\begin{abstract}

The effect of surface roughness on the structure of liquid
crystalline fluids near solid substrates is studied by Monte Carlo
simulations. The liquid crystal is modelled as a fluid of soft ellipsoidal
molecules and the substrate is modelled as a hard wall that excludes
the centres of mass of the fluid molecules. Surface roughness is
introduced by embedding a number of molecules with random positions
and orientations within the wall. It is found that the density and
order near the wall are reduced as the wall becomes rougher (i.e. the
number of embedded molecules is increased). Anchoring coefficients are
determined from fluctuations in the reciprocal space order tensor. It
is found that the anchoring strength decreases with increasing surface
roughness.

\end{abstract}

\maketitle

\section{Introduction}
\label{sec:intro}

The interaction between liquid crystalline (LC) fluids and solid surfaces
has attracted much interest \cite{jerome1992a}. The presence of the
surface breaks the symmetry of the LC phase. As well as being
intrinsically interesting this is technologically important - many
applications of liquid crystals depend on the interaction between the
fluid and an external field, strongly influenced by coupling with
external surfaces.

Most previous studies of LC surface anchoring have assumed that the
surface is homogenous. Two models are commonly used. In the first the
wall is modelled by a perfect crystalline array\cite{schoen1998a}. The
second, more coarse grained model, uses an external potential function
that depends only on the distance from the wall
\cite{cleaver1997a,cheung2004a}. While attractive from a theoretical
standpoint, it has long been recognised that deviations from these
ideal surfaces can affect the properties of the surface \cite{langmuir1918a}.
One notable example of this is the reduction of the order parameter of
nematic liquid crystals at SiO
surfaces\cite{barberi1990a,martinot-lagarde1996a}. This contrasts with
measurements made on other surfaces\cite{jerome1992a} (e.g. rubbed
polyimide) and with most simulation and theoretical studies that give
a higher order parameter at the LC-solid interface. Electron
micrographs show that SiO surfaces are extremely 
rough\cite{martinot-lagarde1997a}, which gives rise to the disordering 
effect of the surface.

In this paper the structures of nematic and isotropic fluids near
rough walls are studied. The effect of roughness is incorporated by
embedding a number of molecules in an otherwise smooth wall. These
are placed and orientated randomly. Similar models have been used for simple
fluids \cite{tang1995a,dong1999a} and it is hoped that this simple
model may give insights into the behaviour of molecular fluids near
rough or porous surfaces. Two aspects of the effect of the surface
roughness on the LC fluid are studied. Firstly the change in the
structure of the fluid was examined. Secondly the effect of surface
roughness on the anchoring properties of the LC. The contribution of
this surface anchoring to the free energy is often taken to be of the
Rapini-Popoular form\cite{rapini1969a}
\begin{equation}\label{eqn:rapinipoplar}
  F_{surf}=W \sin^2(\theta-\theta_0)
\end{equation}
where $\theta-\theta_0$ is the angle between the director at the
surface and the surface's 'easy-axis'. $W$ is the surface anchoring 
coefficient. This depends on both the properties of the bulk liquid
crystal and on the interaction between the liquid crystal and the
surface, so may be expected to vary with surface roughness. As this is
a key property in applications of liquid crystals it would be
interesting to see how this is affected by changes in the surface morphology.

This paper is organised as follows. Details of the simulation,
including the method used for calculating the anchoring coefficient,
are given in the next section. The structure of the fluid confined
between rough walls is given in Sec. \ref{sec:structure} while results
for the anchoring coefficient are presented in
Sec. \ref{sec:anchor}. Finally some concluding remarks are given in
Sec. \ref{sec:summary}.

\section{Simulation}
\label{sec:sim}

\subsection{Simulated Systems}

In order to simulate large systems, a simple intermolecular
potential is used. This models the fluid as a system of
soft ellipsoidal molecules interacting through a simplified version of
the popular Gay-Berne (GB) potential \cite{gay1981a}. In particular
this has two major simplifications. First the
orientation dependence of the energy parameter is suppressed.
Secondly the potential is cut off and shifted at the potential
minima. These changes lead to a much simplier phase diagram than
the GB potential, showing only nematic and isotropic phases, closer to
the phase behaviour of the hard ellipsoid \cite{frenkel1985a} or hard
gaussian overlap \cite{demiguel2001a}  potentials. This potential is also more
computationally efficient than the full GB potential.

The interaction between two molecules $i$ and $j$, with positions
$\mathbf{r}_i$ and $\mathbf{r}_j$, and orientations
$\mathbf{u}_i$ and $\mathbf{u}_j$ is given by
\begin{equation}\label{eqn:vse1}
  V(\mathbf{r}_{ij},\mathbf{u}_i,\mathbf{u}_j)= 
  \left\{ \begin{array}{c @{,\:\:} c}
  4\epsilon_0\left[\rho^{-12}-
    \rho^{-6}\right]+1
  &
  \rho \le 2^{1/6}
  \\
  0
  &
  \mbox{otherwise}
  \end{array}
  \right.
\end{equation}
where $\epsilon_0$ is the energy unit, 
$\mathbf{r}_{ij}=\mathbf{r}_i-\mathbf{r}_j$, and
\begin{equation}
  \rho(\mathbf{r}_{ij},\mathbf{u}_i,\mathbf{u}_j)=
  \frac{r_{ij}-\sigma(\hat{\mathbf{r}}_{ij},\mathbf{u}_i,\mathbf{u}_j)+\sigma_0}
  {\sigma_0}.
\end{equation}
$r_{ij}=|\mathbf{r}_{ij}|$,
$\hat{\mathbf{r}}_{ij}=\mathbf{r}_{ij}/r_{ij}$, and  $\sigma_0$ is the
$\sigma(\hat{\mathbf{r}}_{ij},\mathbf{u}_i,\mathbf{u}_j)$ is
the shape function given by \cite{berne1975a}
\begin{eqnarray}\label{eqn:vse3}
  \sigma(\hat{\mathbf{r}}_{ij},\mathbf{u}_i,\mathbf{u}_j) &=&
  \sigma_0
  \left\{1-\frac{\chi}{2}\left[
    \frac{(\hat{\mathbf{r}}_{ij}.\mathbf{u}_i+
           \hat{\mathbf{r}}_{ij}.\mathbf{u}_j)^2}
         {1+\chi\mathbf{u}_i.\mathbf{u}_j}\right.\right.
         \nonumber\\
        & &+\left.\left.
    \frac{(\hat{\mathbf{r}}_{ij}.\mathbf{u}_i-
           \hat{\mathbf{r}}_{ij}.\mathbf{u}_j)^2}
         {1-\chi\mathbf{u}_i.\mathbf{u}_j}\right]\right\}^{-1/2}.
\end{eqnarray}
This approximates the contact distance between two ellipsoids. In \eref{vse3}
$\chi=(\kappa^2-1)/(\kappa^2+1)$ is the anisotropy parameter, where
$\kappa$ is the elongation (for the molecules studied here $\kappa=3$).

\begin{figure}
\includegraphics[width=7cm]{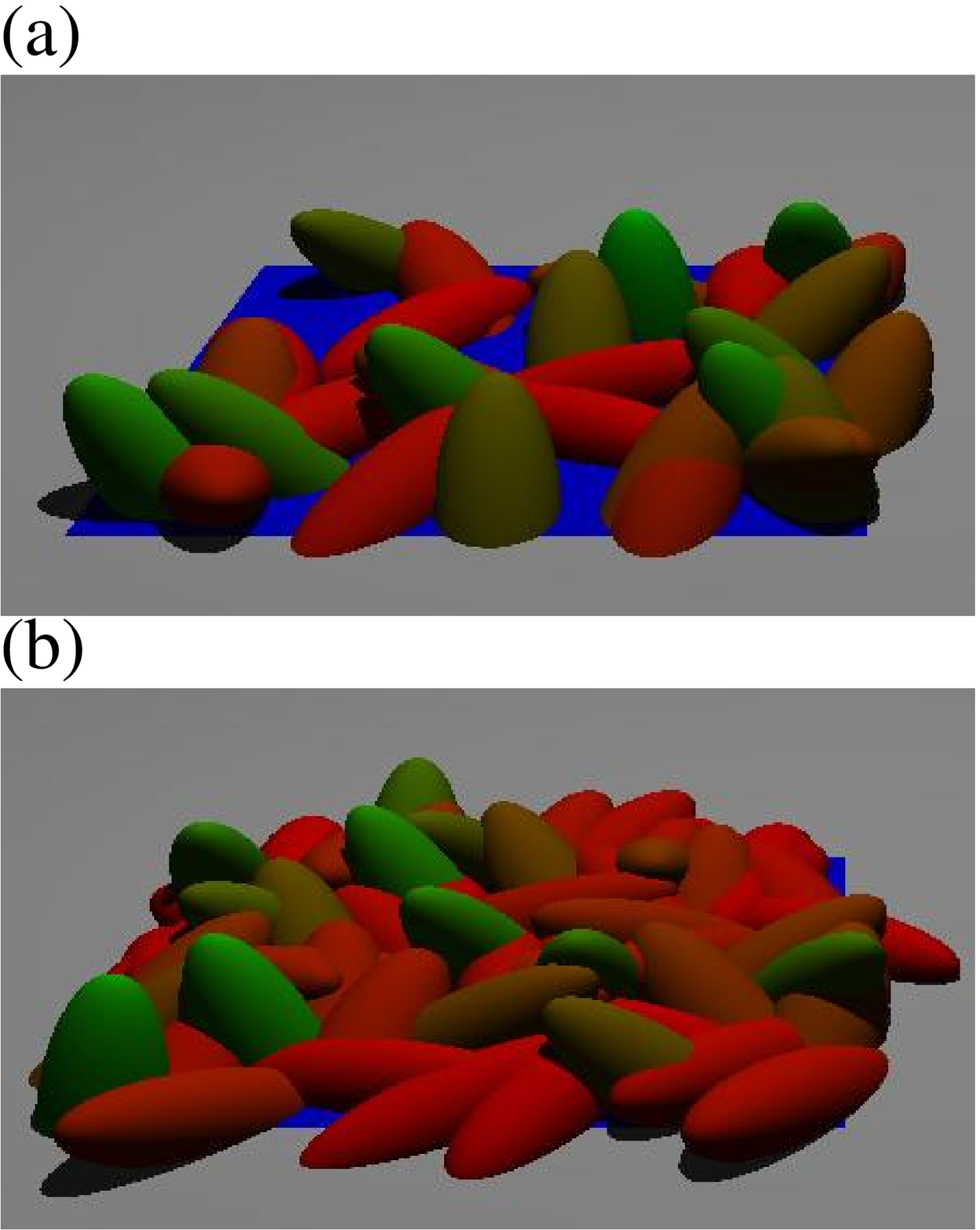}
\caption{
\label{fig:rw_pix}
(Color online) Example rough wall configurations for (a) $\Sigma=0.2$
and (b) $\Sigma=0.4$.
}
\end{figure}

The wall is represented by a hard core potential acting upon the
centres of mass of the molecules. Previous studies have shown that
this gives rise to homeotropic alignment at the wall
\cite{allen1999a}. 
Roughness is introduced by embedding a number of molecules, $N_w$ in
the wall. These were given random positions and orientations which
were kept fixed during the simulations. While generating these surface
configurations interactions between the surface molecules were
ignored, thus these molecules may overlap. It should be noted that
these molecules do not correspond to {\it real} molecules, rather they
are used as a convient way of introducing
inhomogenity into the wall.The roughness of the wall was
characterised by the surface density of these embedded molecules
$\Sigma=N_w/A$.
Some example wall structures are shown in Fig.
\ref{fig:rw_pix}. To ensure some sampling of surface configurations
three different surfaces were studied for each pair of $\rho$ and $\Sigma$.

Simulations were performed at two average densities, $\rho=0.314$ and
$\rho=0.30$. For the higher density the fluid confined between smooth
walls was nematic, while it is isotropic for the lower density. The
simulated systems were composed of 1200 fluid molecules and up to 63
molecules embedded in each wall. Throughout this reduced units defined
by the molecular width $\sigma_0$ and the energy unit $\epsilon_0$ are
used. A reduced temperature of 0.5 was used for both densities.

\subsection{Simulation observables}

The orientational order may be characterised by the usual nematic order
parameter. This is given by the largest eigenvalue of the ordering
tensor, defined as
\begin{equation}\label{eqn:ordert}
  Q_{\alpha\beta}=\frac{1}{N}\sum_{i=1}^{N} 
  \left(\frac{3}{2}u_{i\alpha}u_{i\beta}-\frac{1}{2}\delta_{\alpha\beta}
  \right), ~~ \alpha,\beta = x,y,z
\end{equation}
where $\mathbf{u}_i$ is the orientation of the $i$th molecule and
$\delta_{\alpha\beta}$ is the Kronecker delta function. It may also be
informative to consider the order parameter in the cell bulk and near
the surface, $S^{bulk}$ and $S^{surf}$. $S^{bulk}$  is
calculated for molecules within the region $l_z/4 \leq z \leq 3l_z/4$,
while $S^{surf}$ is calculated for molecules within 1 $\sigma_0$ of
the surface.

The distribution of molecules in the simulation cell can be described
by the density profile $\rho(z)$. To describe the ordering through the
cell, the ordering tensor \eref{ordert} can be calculated throughout
the cell. Diagonalising this gives the order parameter profiles
($q_+(z)$, $q_0(z)$, and $q_-(z)$). These can be expressed as $S(z)$,
$S(z)+\frac{1}{2}S_{xy}(z)$, and $S(z)-\frac{1}{2}S_{xy}(z)$, where
$S(z)$ is the nematic order parameter and $S_{xy}(z)$ is the
biaxiality parameter. 

The nematic director $\mathbf{n}(z)$ can be identified with the
eigenvector of the ordering tensor corresponding to the largest eigenvalue. 

While the presence of layers may be deduced from the density profiles
it may be useful to quantify the degree of translational order. The
smectic order parameter may be introduced for this purpose 
\cite{steuer2004a,luckhurst1999a}. This is given by
\begin{equation}\label{eqn:sm_op}
  \rho_1 = \left<\left| \frac{1}{N} \sum_{j=1}^{N} 
  \exp\left(\frac{2\pi iz_j}{d}\right)
  \right|\right>,
\end{equation}
where $d$ is the layer periodicity. This is initially unknown and is
take to be the value that maximises $\rho_1$ \cite{luckhurst1999a}.

\subsection{Director fluctuations and surface anchoring}

The surface anchoring coefficient is determined by the director
fluctuation method \cite{allen2000a,allen2002a}. This method relates
thermal director fluctuations in a confined geometry to the zenithal
anchoring coefficient, in a similar manner as the fluctuations in a
bulk LC can be related to the bulk elastic constants 
\cite{forster,allen1996a}. The theory for this has been extensively 
developed elsewhere and this section will contain only the briefest of 
outlines.

As for the bulk elastic constants the zenithal anchoring coefficient
may be determined by fitting elastic theory predictions of
fluctuations in the ordering tensor to those determined from
simulations. The reciprocal space ordering tensor is given by
\begin{equation}\label{eqn:ordert_k}
  Q_{\alpha\beta}(\mathbf{k}) = 
  \frac{V}{N}\sum_j Q_{\alpha\beta}^j
  \exp\left(i\mathbf{k}.\mathbf{r}_i\right).
\end{equation}
Fluctuations can be calculated from simulation
\begin{eqnarray}
  \left<\left|Q_{\alpha\beta}(k_z)\right|^2\right> &=& 
  \frac{V^2}{N^2}\left[
    \left(\sum_j Q_{\alpha\beta}^j \cos(k_zz_j)\right)^2 \right.\nonumber\\
    & & +
    \left.\left(\sum_j Q_{\alpha\beta}^j \sin(k_zz_j)\right)^2\right]
    \label{eqn:ordert_fl},
\end{eqnarray}
The corresponding elastic theory predicts that there fluctuations are given by
\cite{allen2000a}
\begin{eqnarray}
  \left<\left|Q_{\alpha\beta}(k_z)\right|^2\right> &=& 
  \frac{9}{8}k_BT\frac{S^2V}{K_{33}}
  \sum_{q_z}
  \frac{\chi^2+\zeta^2}{q_z^2(2\zeta+\zeta^2+\chi^2)}\times 
  \nonumber\\
  & &
  \left|
  \frac{e^{i(\kappa+\chi)}-1}{\kappa+\chi}+
  \left(\frac{i\chi-\zeta}{i\chi+\zeta}\right)
  \frac{e^{i(\kappa-\chi)}-1}{\kappa-\chi}\right|^2 \nonumber\\
  & &
  \label{eqn:ordert_fl_elastic}
\end{eqnarray}
where $K_{33}$ is the bend elastic constant. $q_z$ is a wave vector
with a discrete spectrum\cite{allen2000a}, $\chi=q_zL_e$, and $\kappa=k_zL_e$. $\zeta$ is
the anchoring strength parameter 
\begin{equation}
  \zeta = \frac{WL_e}{K_{33}} = \frac{L_e}{\lambda}
\end{equation}
where $W$ is the zenithal anchoring coefficient and $\lambda$ is the
extrapolation length. $L_e$ is the cell thickness appearing in the
elastic theory; this is not necessarily equal to the simulation cell
thickness, $L_z$. In fitting the elastic theory to simulation
profiles $L_e$ and $\zeta$ are the fitting parameters. $K_{33}$ has been
determined from simulation for a nearby state point\cite{schmid2001a} 
($\rho=0.30$). While this value ($K_{33}=1.48$) is likely to be too
large for some of the systems studied here, this should be sufficient for a
qualitative study.

\section{Fluid Structure}
\label{sec:structure}

\subsection{High Density Fluid}

\begin{figure}
\includegraphics[width=4.2cm]{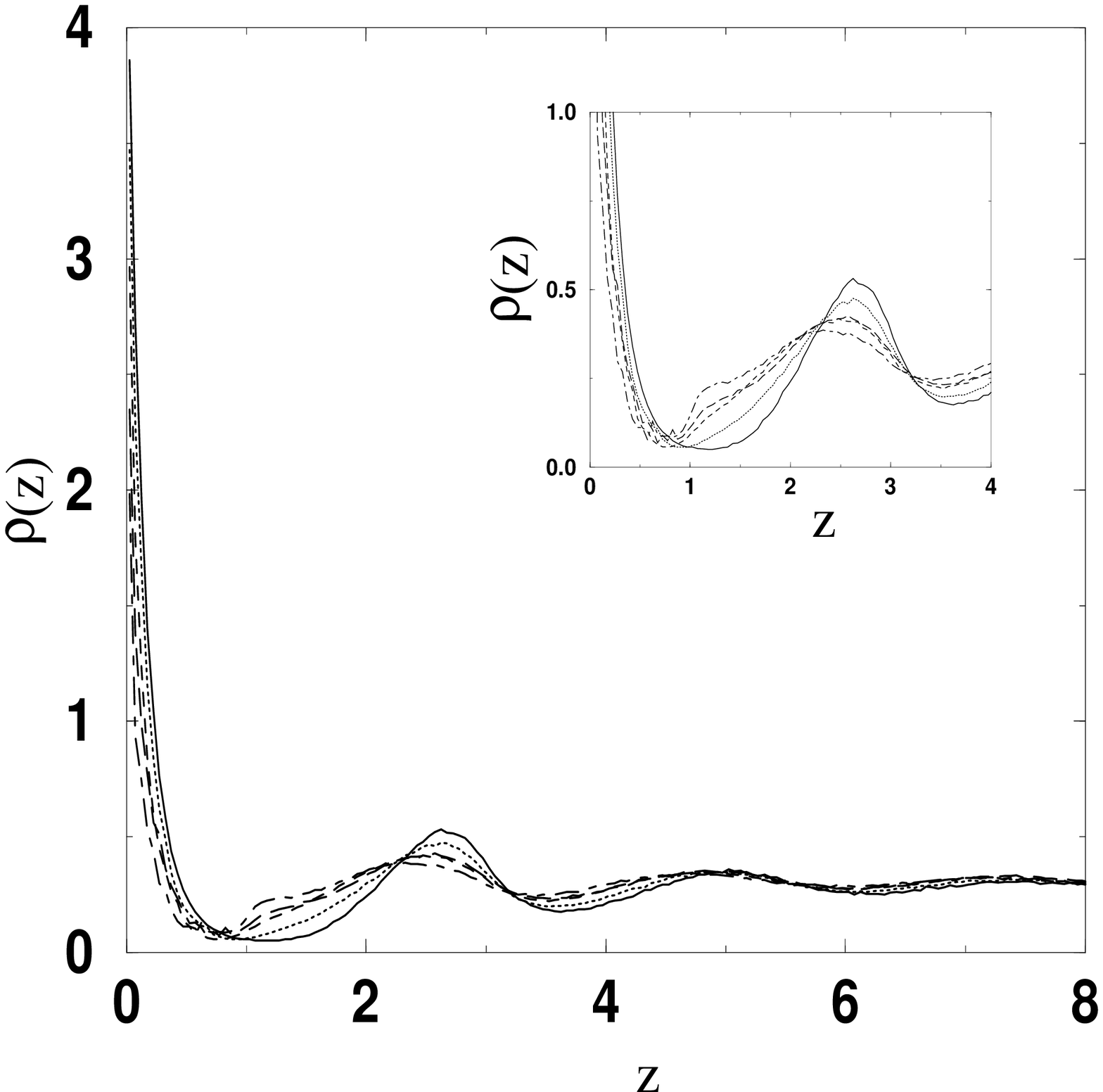}
\includegraphics[width=4.2cm]{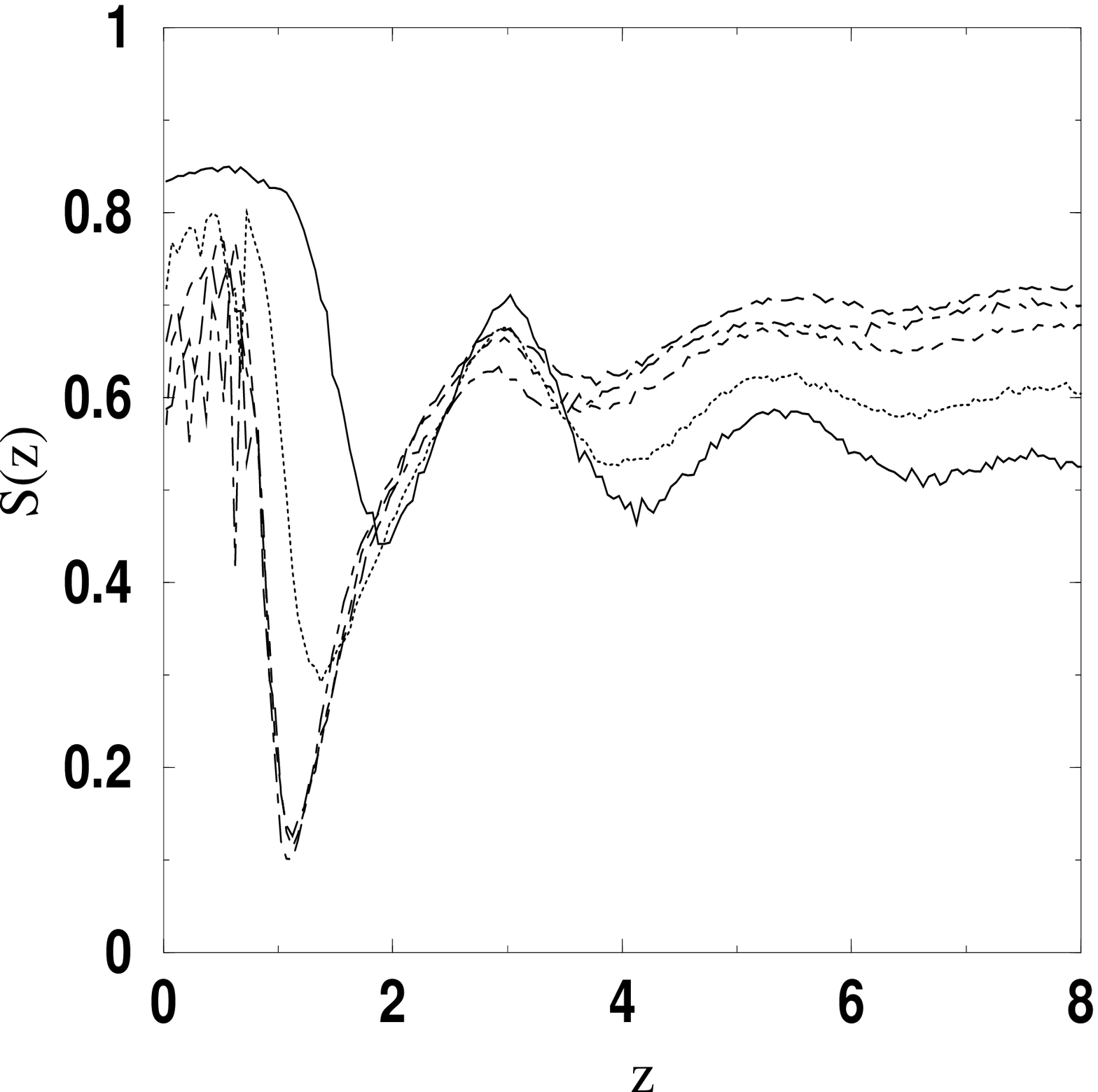}
\includegraphics[width=4.2cm]{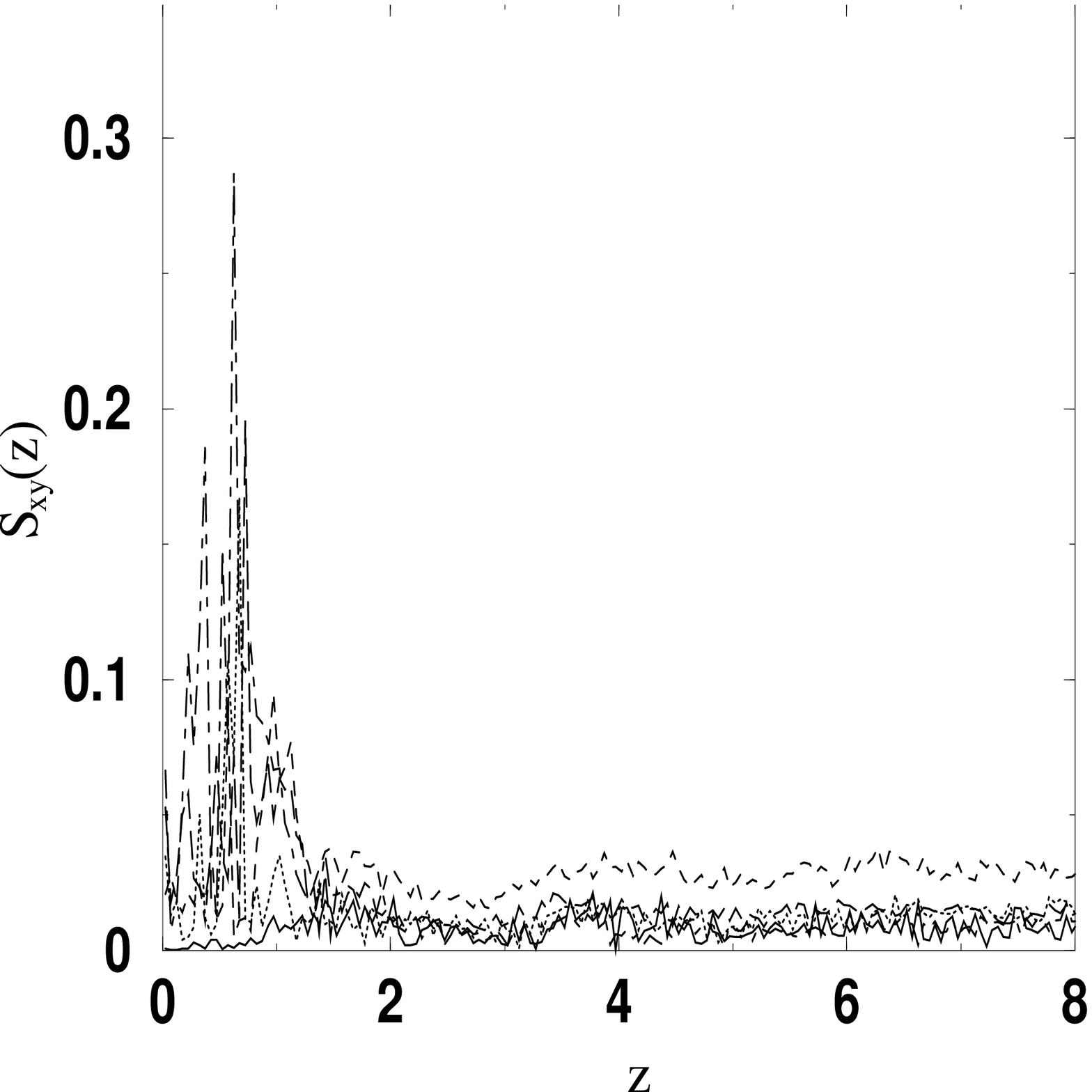}
\includegraphics[width=4.2cm]{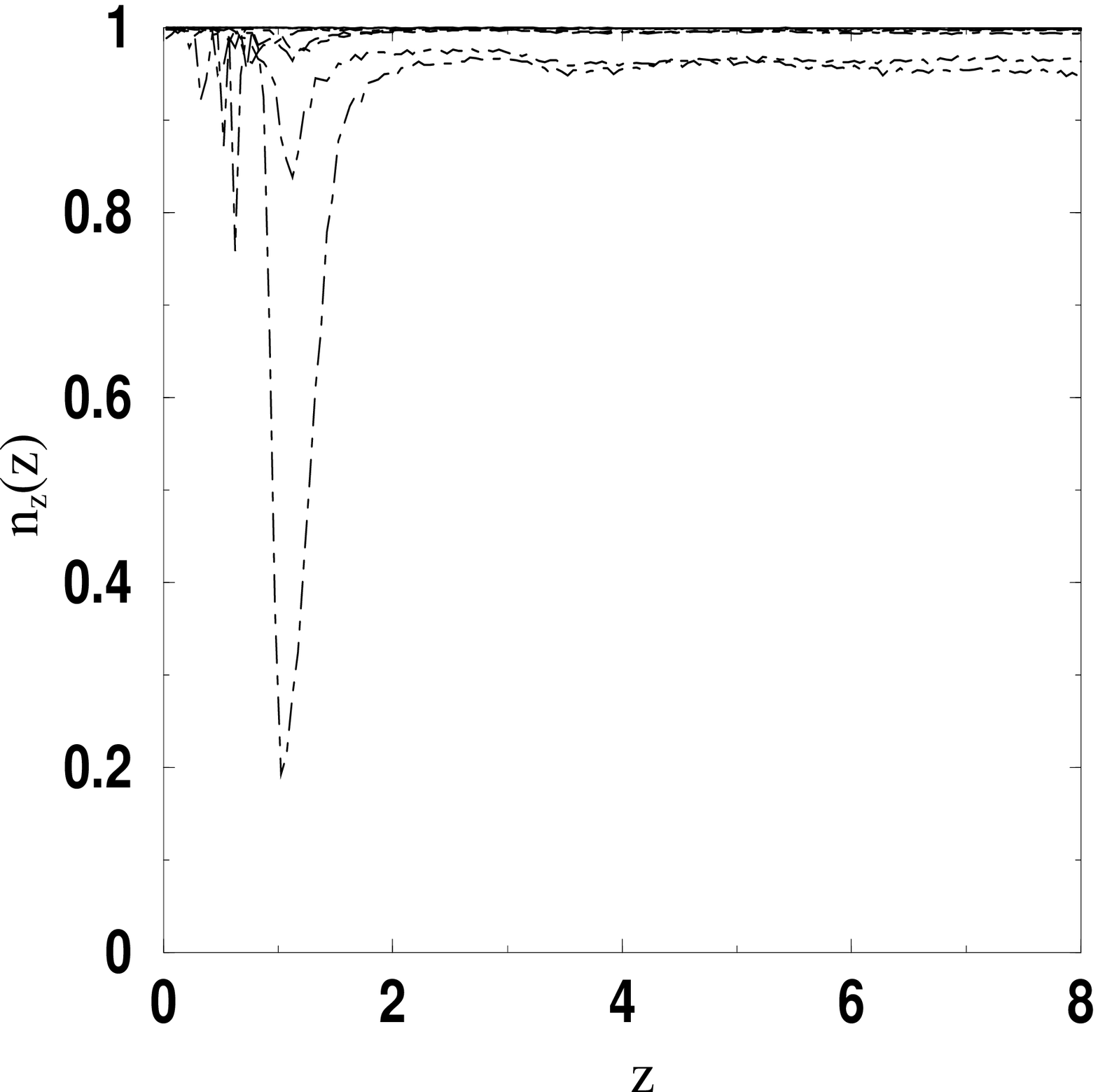}
\caption{
\label{fig:rho0.314}
(a) Density profiles for the high density  fluid near
rough walls. The density profile for grafting density $\Sigma=0$ is
shown by the solid line, for $\Sigma=0.1$ dotted line, $\Sigma=0.2$
dashed line, $\Sigma=0.3$ long dashed line, and $\Sigma=0.4$ the
dashed dotted line. Inset shows the density profiles around the
minima. Symbols as in main figure. \\
(b) Order parameter profiles for high density fluid near rough walls. 
Symbols as in (a). \\
(c) Biaxiality ($S_{xy}$) profiles for high density fluid near rough
wall. Symbols as in (a). \\
(d) $z$ component of the director for the high density fluid. Symbols
as in (a). 3 $n_z(z)$ profiles are shown for $\Sigma=0.4$.
}
\end{figure}

The density profiles for the high density fluid are shown in Fig. 
\ref{fig:rho0.314}a. The effect of the wall roughness is most
apparent near the wall. Here the density near the surface decreases
with increasing $\Sigma$. This is caused by the decrease in available 
volume near the wall due to the embedded molecules. Values of
the density near the wall are presented in Tab. \ref{tab:effdens}. The 
surface density falls from 0.72 for the smooth wall to 0.34 for the
rough wall with $\Sigma=0.4$.

Another noticeable difference is that the second peak (at $z=2.8$ for
the plain wall) becomes
broader. This arises from the surface disorder disturbing the layer
structure and has been observed in simulations of Lennard-Jones fluids 
\cite{dong1999a}. This can more clearly be seen in the inset, which
shows the detail of the density
profiles around the minima. The disruption of the translational
ordering caused by the embedded molecules can be seen by considering
the smectic order parameter (\eref{sm_op}). Values for these are
presented in Tab. \ref{tab:effdens}. As can be seen $\rho_1$ markedly 
decreases with increasing grafting density, as would be expected for
increasing translational disorder.

Far from the wall the profiles all tend to a
constant values, indicating a layer of bulk fluid. The density of this
layer increases slightly with increasing grafting density. This
arises as the embedded molecules exclude fluid molecules from regions
near the wall, increasing the number of molecules in the cell
bulk. This is a consequence of having a fixed cell size and may be
avoided by using $NpT$ simulations. Quantitively this can be seen by
examining the densities in the
cell bulk. Values for this are presented in Tab. 
\ref{tab:effdens}. The density in the bulk of the cell goes from 0.29
for the smooth wall to 0.31 for the highest grafting densities.

Figure \ref{fig:rho0.314}(b) shows the order parameter profiles for
different values of $\Sigma$. As can be seen the value of the order
parameter at the wall is lower for higher grafting densities. This 
is caused by the disorientating effect of the embedded molecules. This
disorientating effect also leads to a deeper minima. For $\Sigma \geq
0.2$ this leads to a small layer (approximately 1 molecular width
thick) of almost
isotropic fluid. The position of this minima moves closer to the wall
with increasing surface roughness. For the smooth wall this minima is
at approximately $z=2$, while for the highest grafting densities it appears at
about $z=1.1$. Again this is attributable to the disruption in the
surface induced layering. As for $\rho(z)$ the second peak becomes
broader with increasing $\Sigma$. Finally, as can be seen from Tab. 
\ref{tab:effdens} the bulk order
parameter $S^{bulk}$ increases with increasing $\Sigma$. This is a
consequence of the increasing density in the centre of the cell due to
the excluded volume effect of the embedded molecules.  It is
noticeable that for $\Sigma \geq 0.2$ $S^{bulk}$ becomes larger than 
$S^{surf}$. 

The biaxiality profiles are shown in Fig. \ref{fig:rho0.314}(c). For
the smooth wall the this is essentially zero (the largest value
is 0.04) reflecting the cylindrical symmetry around the $z$ axis. However, for
the rough walls there is are sizable peaks in the biaxiality
profiles. These are stronger for larger values of $\Sigma$ and are in
the region of $0.5 \leq z \leq 1.3$, corresponding to regions of low
order. This surface induced biaxiality
has been seen for simulations of LCs near grooved surfaces \cite{stelzer1997a}.

\begin{Table}
{\label{tab:effdens}Densities and order parameters for the simulated
  systems. $\rho_{bulk}$ and $\rho_{surf}$ are the bulk and surface
  densities, $S$, $S^{bulk}$ and $S^{surf}$ are the total, bulk and surface
  order parameters, and $\rho_1$ is the smectic order
  parameter. Errors in the last decimal place are in parenthesises.
}
\begin{tabular}{| c | c | c | c | c | c | c | c |}
\hline
$\rho$ & $\Sigma$ & $\rho_{surf}$ & $\rho_{bulk}$ & $S$ & $S^{surf}$  & 
$S^{bulk}$ & $\rho_1$ \\[0.5ex]
\hline
0.314 &  0  & 0.72(1) & 0.286(3)  &  0.60(3)   &  0.84(1)   &  0.53(6) &  
0.14(2)  \\
0.314 & 0.1 & 0.61(1) & 0.294(2)  &  0.64(2)   &  0.76(2)   &  0.65(3) &  
0.12(2)  \\
0.314 & 0.2 & 0.48(3) & 0.304(3)  &  0.67(2)   &  0.65(3)   &  0.72(2) &  
0.10(2)  \\
0.314 & 0.3 & 0.43(5) & 0.308(4)  &  0.69(2)   &  0.65(9)   &  0.75(2) &  
0.09(2)  \\
0.314 & 0.4 & 0.33(2) & 0.307(9)  &  0.66(8)   &  0.58(7)   &  0.73(7) &  
0.07(2)  \\
\hline
0.300 &  0  & 0.70(1) & 0.273(2)  &  0.28(3)   &  0.81(1)   &  0.10(4) &  
0.14(2)  \\
0.300 & 0.1 & 0.59(1) & 0.281(2)  &  0.34(7)   &  0.72(3)   &  0.27(9) &  
0.12(2)  \\
0.300 & 0.2 & 0.46(2) & 0.291(3)  &  0.52(6)   &  0.61(3)   &  0.59(5) &  
0.10(2)  \\
0.300 & 0.3 & 0.41(2) & 0.293(3)  &  0.51(4)   &  0.53(9)   &  0.60(4) &  
0.09(2)  \\
0.300 & 0.4 & 0.38(1) & 0.300(3)  &  0.59(6)   &  0.54(5)   &  0.69(3) &  
0.07(2)  \\
\hline
\end{tabular}
\end{Table} 

Figure \ref{fig:rho0.314}(d) shows the $z$ component of the director
for each $\Sigma$. In the cell bulk the director is essentially parallel
to the $z$ axis. For $\Sigma \geq 0.2$ there is a tilt away
from the $z$ axis at about the position of the order minima. As may be
expected this is most pronounced for the $\Sigma=0.4$ wall. In Fig. 
\ref{fig:rho0.314}d the profiles for each of the $\Sigma=0.4$ walls
are shown separately. It can be seen that the size of this tilt differs
strongly for different wall configurations (for the largest the tile
angle is approximately $79^\circ$). For the larger
tilt angles this propagates into the bulk of the fluid leading to a
director tilted up to $16^\circ$ from the $z$-axis. It is not clear
how a randomly generated wall gives rise to a titled configuration in
the bulk. Similar behaviour has been seen in a recent study of a LC
near a planar wall with perpendicularly grafted
rods\cite{allen2004a}.  In that case the bulk tilt was caused by the 
competition between the wall (which promoted planar alignment) and the 
embedded molecules. As it appears only for a subset of the walls
studied here it would be desirable to consider further wall configurations.

\begin{figure}
\includegraphics[width=6cm]{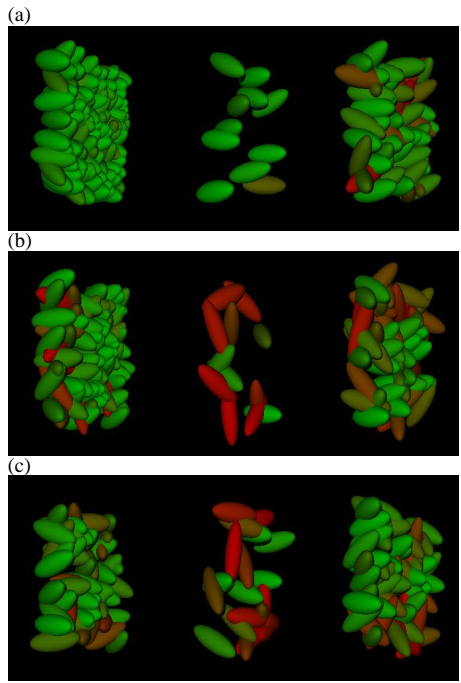}
\caption{
\label{fig:rho0.314_snp}
(Color online) Simulation configurations showing molecules within 2.5
$\sigma_0$ of the surface for (a) $\Sigma=0$, (b) $\Sigma=0.2$, and (c)
$\Sigma=0.4$. For each $\Sigma$ the left most picture shows molecules
with $0 \leq z \leq 0.5$, the centre pictures shows $0.5 \leq z \leq
1.5$, and the right most shows $1.5 \leq z \leq 2.5$.
}
\end{figure}

The previous discussion may be illuminated by examination of
simulation configurations. Figure \ref{fig:rho0.314_snp} shows 
configurations of for
$\Sigma=0.0$ (smooth wall), $\Sigma=0.2$, and $\Sigma=0.4$. The
disordering effect of the rough wall can be seen in the first and
second layers (left and right most pictures). However, the fluid
between these two layers shows the most noticeable change with
increasing $\Sigma$. For the smooth wall the molecules in this region
are still well ordered parallel to the $z$-axis. With increasing
$\Sigma$ the molecules in this region become increasingly
disordered. This gives rise to the deeper minima seen in the order
parameter profile (Fig. \ref{fig:rho0.314}(b)). Additionally it can be
seen that many of the molecules lie in the $xy$ plane, giving rise to
the biaxiality peak and the tilt of the director away from the $z$
axis. This behaviour is similar to that seen in
simulations of smectic liquid crystals \cite{frenkel1995a} where
molecules in the region between the layers are seen to align either
parallel or normal to the layers. These planar oriented molecules
possibly give rise to the bulk tilt seen in some cases.
Finally the number of molecules in
this region visibly increases with $\Sigma$.

\subsection{Low Density Fluid}

Here the density and order parameter profiles for the low density
system are discussed. For the smooth wall the density in the bulk of
the cell is 0.27 (Tab. \ref{tab:effdens}), just below the isotropic-nematic transition density
for this system ($\rho_{I-N}=0.287$). As the density in the cell bulk
increases with $\Sigma$, for $\Sigma \geq 0.2$ the fluid in the cell
bulk is nematic rather than isotropic.

\begin{figure}
\includegraphics[width=4cm]{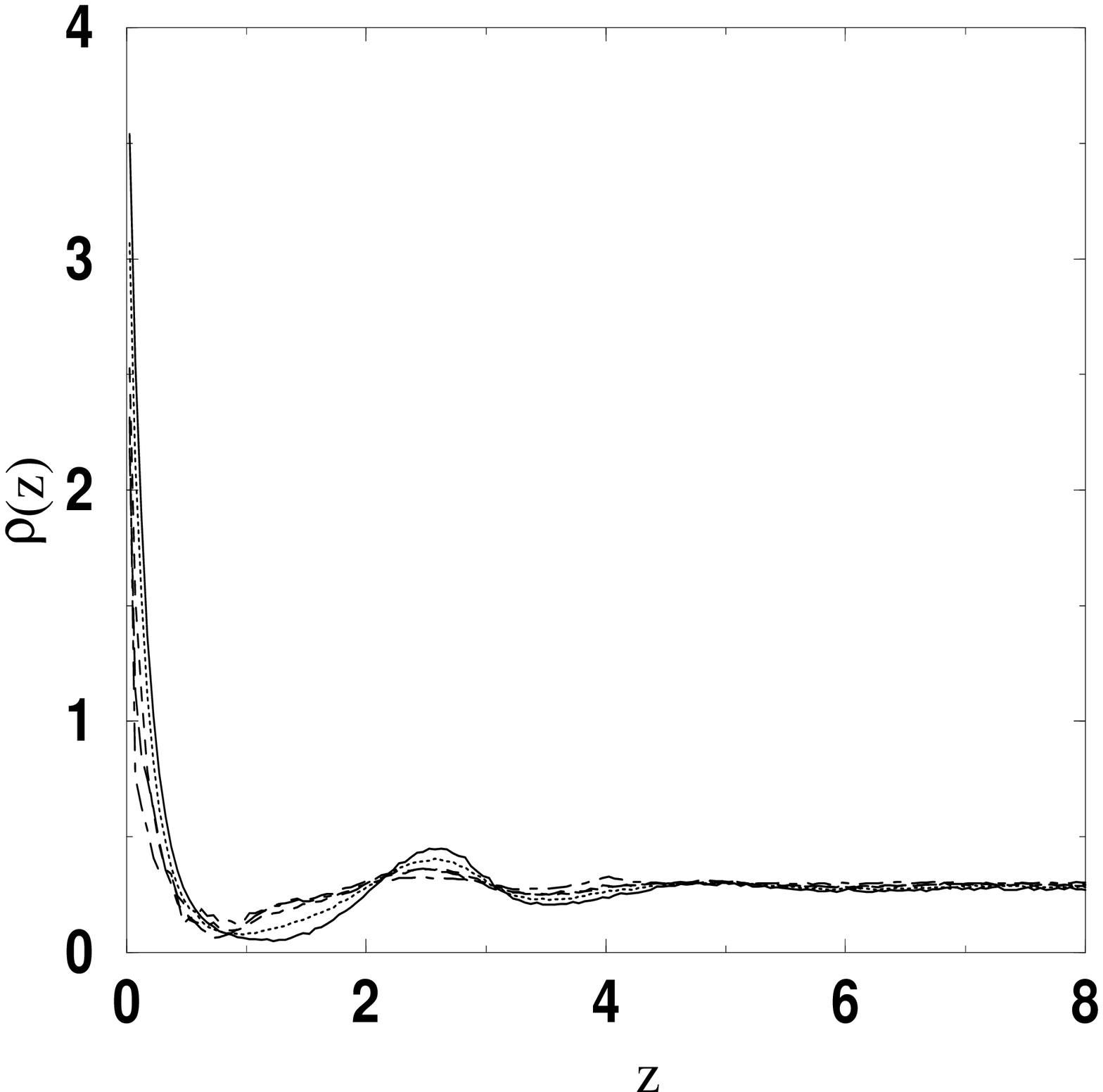}
\includegraphics[width=4cm]{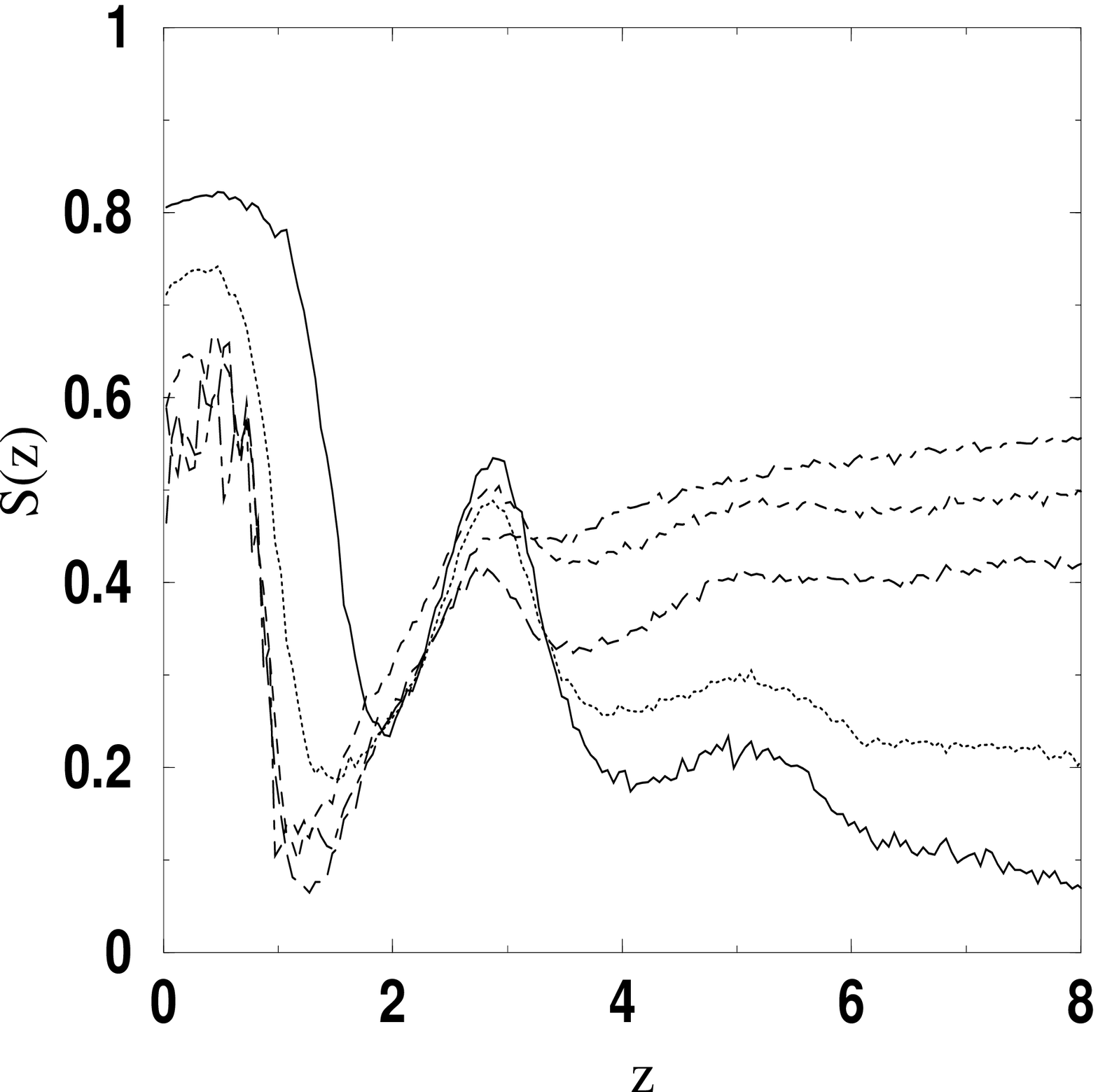}
\caption{
\label{fig:rho0.30}
(a) Density profiles for the low density fluid near
rough walls. The density profile for grafting density $\Sigma=0$ is
shown by the solid line, for $\Sigma=0.1$ dotted line, $\Sigma=0.2$
dashed line, $\Sigma=0.3$ long dashed line, and $\Sigma=0.4$ the
dashed dotted line. \\
(b) Order parameter profiles for the low density fluid. Symbols as in
(a). 
}
\end{figure}

The density profiles for the low density fluid are shown in
Fig. \ref{fig:rho0.30}(a). The changes in the density profile with
increasing $\Sigma$ are similar to those in the high density system -
the density at contact decreases with $\Sigma$ and the second peak becomes more
diffuse. Again this can be gleaned from the decrease in the value of
the smectic order parameter with $\Sigma$ (Tab. \ref{tab:effdens}).
It is interesting to note that the values of $\rho_1$ obtained in this
system are very similar to those for the higher density system,
indicating the similarity in the structure of both systems.

Shown in Fig. \ref{fig:rho0.30}(b) are the order parameter
profiles. As in the high density fluid the value of the order parameter at
the wall decreases as $\Sigma$ increases. The order parameter profile
also shows a deeper minima with
increasing surface roughness. It is noticeable that even in this lower
density case there is not an appreciable layer of isotropic fluid
between the wall and bulk fluid. This has been predicted to happen near
rough walls as a
consequence of the competition between the bulk director and the local
boundary conditions \cite{sluckin1995a}.

\section{Surface Anchoring}
\label{sec:anchor}

\begin{figure}
\includegraphics[width=4cm]{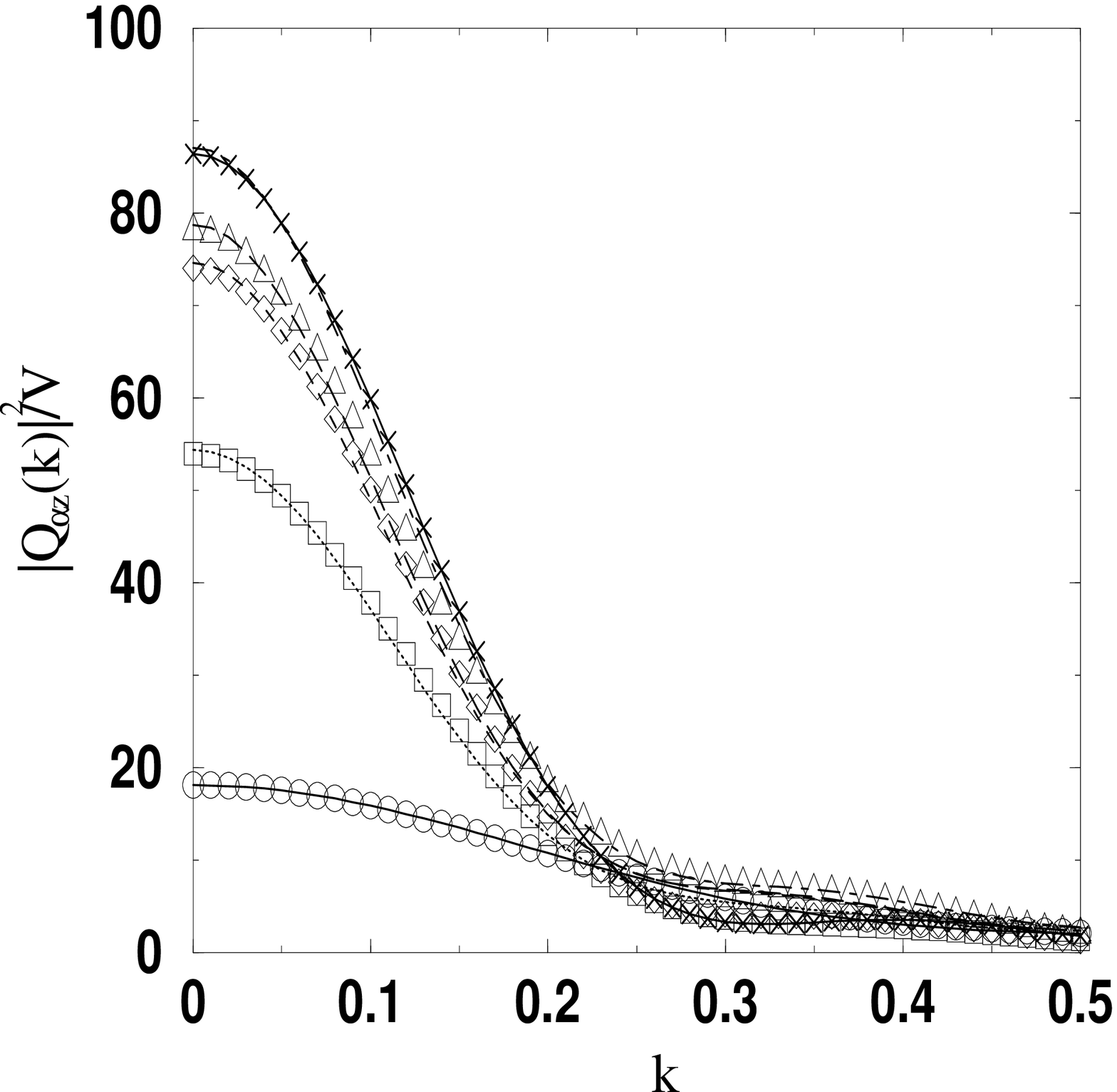}
\includegraphics[width=4cm]{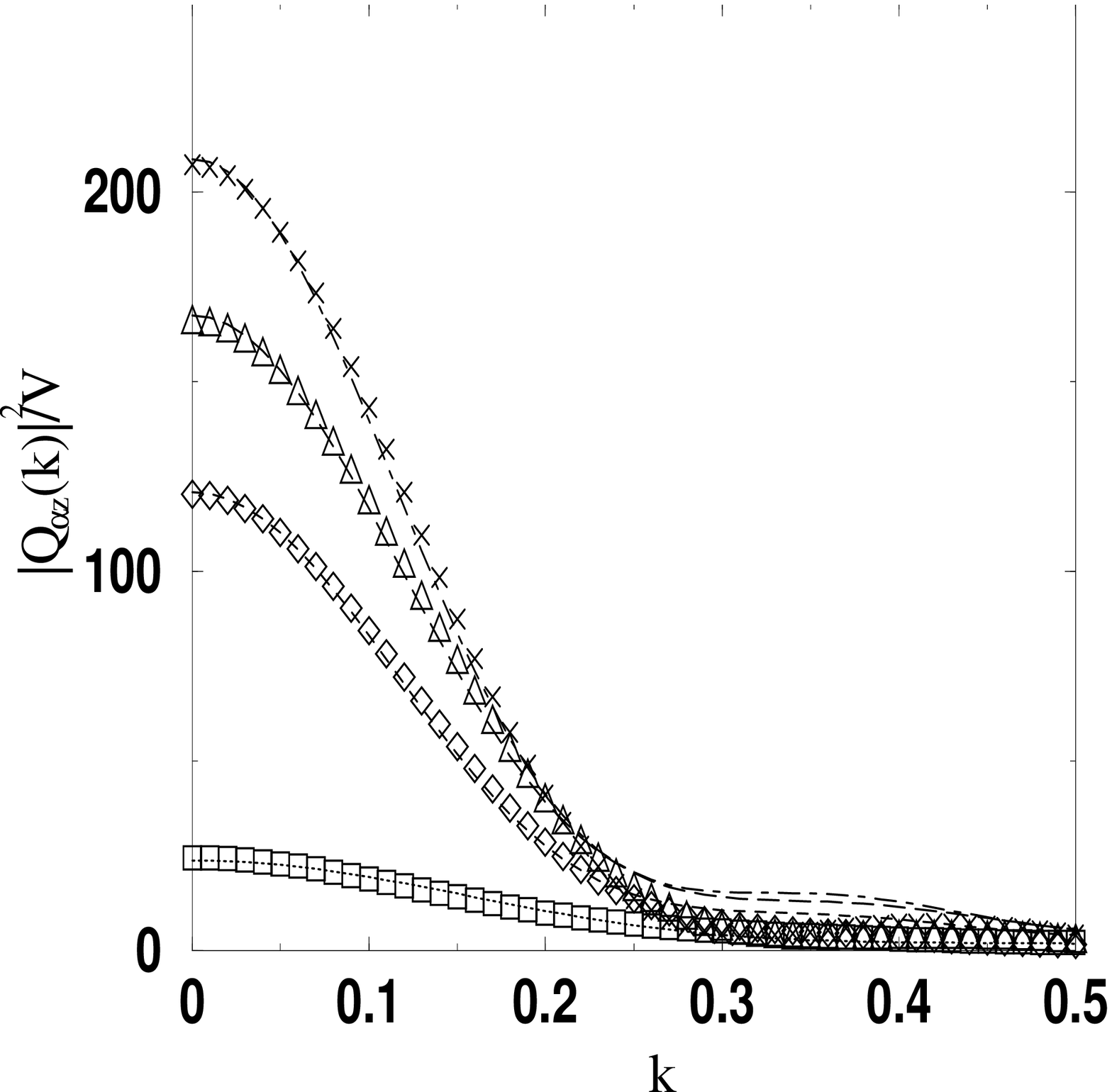}
\caption{
\label{fig:ordert_fl}
Order tensor fluctuations (normalised by cell volume) as a function of 
wavevector for (a) high density and (b) low density fluids. In both
graphs the simulation data
is denoted by symbols (circles $\Sigma=0$, squares $\Sigma=0.1$,
diamonds $\Sigma=0.2$, triangles $\Sigma=0.3$, and crosses
$\Sigma=0.4$) and the elastic theory data is shown by lines
(continuous line $\Sigma=0$, dotted line $\Sigma=0.1$, dashed line
$\Sigma=0.2$, long dashed line $\Sigma=0.3$, and dot dashed line
$\Sigma=0.4$). The order tensor fluctuations for $\Sigma=0.0$ are
shown only in (a).
}
\end{figure}

Shown in Fig. \ref{fig:ordert_fl} are the order tensor fluctuations as
a function of wavevector. As can be there is good agreement between the
simulation and elastic theory curves, especially for small $k_z$. 

The fitted values for the anchoring coefficients are given in Tab. 
\ref{tab:anchdat} along with values of the extrapolation length
$\lambda$ and the surface anchoring coefficient $W$. For both bulk 
densities $\zeta$ tends to decrease with increasing $\Sigma$.

\begin{Table}
{\label{tab:anchdat}Fitting data for the order tensor fluctuations
  (Fig. \ref{fig:ordert_fl}). $\zeta$ is the anchoring coefficient,
  and $L_e$ is the elastic theory cell width, which appear in 
  \eref{ordert_fl_elastic}. $\lambda=L_e/\zeta$ is the extrapolation length and
  $W=K_{33}/\lambda$ is the surface anchoring strength.}
\begin{tabular}{| c | c | c | c | c | c | c |}
\hline
$\rho$ & $\Sigma$ & $\zeta$ & $L_{e}$ & $\lambda$ &
  $W$  \\
\hline
0.314  &  0.0  &   5.62  &  16.00  &  2.85  &  0.52  \\
0.314  &  0.1  &   6.51  &  28.39  &  4.36  &  0.34  \\
0.314  &  0.2  &   5.64  &  29.36  &  5.21  &  0.28  \\
0.314  &  0.3  &   4.94  &  26.80  &  5.43  &  0.27  \\
0.314  &  0.4  &   4.55  &  27.78  &  6.11  &  0.24  \\
\hline
0.30   &  0.1  &   3.05  &  18.32  &  6.01  &  0.24  \\
0.30   &  0.2  &   2.88  &  24.60  &  8.54  &  0.17  \\
0.30   &  0.3  &   2.58  &  23.85  &  9.24  &  0.16  \\
0.30   &  0.4  &   2.70  &  25.44  &  9.42  &  0.16  \\
\hline
\end{tabular}
\end{Table}

The behaviour of the elastic theory cell width, $L_{e}$, with
$\Sigma$ deserves comment. For the high density fluid, $L_{e}$
for the smooth wall is 16.00 $\sigma_0$, a few molecular
lengths smaller than the physical cell width ($L_z=24.66$). This is
similar to behaviour seen in previous
simulations\cite{allen2000a,allen2002a} and is due to the formation of
highly ordered layers in the vicinity of the surface. For the rough
walls however, $L_{e}$ is larger than $L_z$. This
increase may be attributable to the rough surface breaking up the
highly ordered surface layer. Thus instead of the elastic theory
boundary conditions being applied at this layer, they are applied
closer to the wall, leading to an increase in $L_e$.

For both bulk densities the extrapolation length increases and the
anchoring coefficient $W$ decreases with $\Sigma$. Thus, as may be
intuitively expected, anchoring on rough surfaces is weaker than on
smooth surfaces.

\section{Summary}
\label{sec:summary}
 
In this paper results of Monte Carlo simulations for a confined fluid
of ellipsoidal molecules have been performed. The effect of surface
roughness on the structure of the fluid has been examined. Roughness
was introduced by embedding a number of molecules, with random
positions and orientations, in otherwise smooth
walls. Increasing the number of molecules embedded in the wall
corresponds to an increasing surface roughness. The simulations were
performed at two bulk densities. For the higher density the fluid in
the bulk of the cell is nematic for the smooth wall case, while for
the lower density it is isotropic.

At both densities studied the effect of increasing surface roughness
is similar. Both the density and order parameter in the region near
the wall decrease as the number of embedded molecules increases. The
decrease in the density arises from the excluded volume of the
embedded molecules, while the decrease in the order can be attributed
to the disorientating effect of the randomly orientated molecules in
the wall. The rough walls also act to smear out the secondary peaks in
the density and order parameter profiles as the embedded molecules
give anchoring points at positions other than at the wall surface. One
side effect of the wall roughness is an increase in the density and
order parameter in the centre of the cell. 

Also studied was the effect of surface roughness on the surface
anchoring strength. For both systems the anchoring was
found to become progressively weaker with increasing surface
roughness.

A number of possible avenues for future work are possible. Calculation
of the anchoring coefficient via alternative
methods\cite{allen1999a,longa1995a} would be useful. As the
formation of a highly ordered layer at the surface is commonly held to
be important for the growth of order in confined liquid crystals, it
may be interesting to investigate the effect
of wall roughness on the phase behaviour of the confined fluid
\cite{steuer2004a}. Integral equation \cite{dong1995a}
or density function theories \cite{schmidt2003a} have been applied to
similar systems of simple fluids and appropriate generalisations
to molecular fluids should also prove useful.

\begin{acknowledgements}

The authors wish to thank the German Science Foundation for
funding. Computational resources for this work were kindly provided by
the Fakult\"{a}t f\"{u}r Physik and  Centre for Biotechnology,
Universit\"{a}t Bielefeld. 

\end{acknowledgements}


\bibliography{paper}

\end{document}